\documentstyle[12pt]{article}
\def\bcn{\begin{center}}
\def\ecn{\end{center}}
\newcommand{\beqn}{\begin{eqnarray}}
\newcommand{\eeqn}{\end{eqnarray}}
\def\spose#1{\hbox to 0pt{#1\hss}}
\def\lsim{\mathrel{\spose{\lower 3pt\hbox{$\mathchar"218$}}
     \raise 2.0pt\hbox{$\mathchar"13C$}}}
\def\gsim{\mathrel{\spose{\lower 3pt\hbox{$\mathchar"218$}}
     \raise 2.0pt\hbox{$\mathchar"13E$}}}
\def\simpropto{\mathrel{\spose{\lower 3pt\hbox{$\mathchar"218$}}
     \raise 2.0pt\hbox{$\propto$}}}
\def\beq{\begin{equation}}
\def\eeq{\end{equation}}
\def\barr{\begin{array}}
\def\earr{\end{array}}
\def\and{\qquad {\rm and } \qquad}
\def\etal{ {\it et al.} }
\def\ie{ {\it i.e.} }

\def\ZPC#1#2#3{{\sl Z.~Phys.} {\bf C#1}, #2 (#3)}
\def\PTP#1#2#3{{\sl Prog. Theor. Phys.} {\bf #1}, #2 (#3)}
\def\PRL#1#2#3{{\sl Phys. Rev. Lett.} {\bf #1}, #2 (#3)}
\def\PRD#1#2#3{{\sl Phys. Rev.} {\bf D#1}, #2 (#3)}
\def\PLB#1#2#3{{\sl Phys. Lett.} {\bf B#1}, #2 (#3)}
\def\PREP#1#2#3{{\sl Phys. Rep.} {\bf #1}, #2 (#3)}
\def\NPB#1#2#3{{\sl Nucl. Phys.} {\bf B#1}, #2 (#3)}

\def\tev{{\rm TeV }}
\def\gev{{\rm GeV }}

\def\mgut{M_{\rm GUT}}
\def\mugut{\mu_{\rm GUT}}
\def\mz{m_{\rm z}}

\def\abs#1{\left|#1\right|}
\def\threebar{\overline{3}}
\def\fivebar{\overline{5}}
\def\tenbar{\overline{10}}
\def\fivteenbar{\overline{15}}
\def\xbar{\overline{x}}
\def\msbar{\overline{\rm MS}}

\def\threetenths{{3\over 10}}
\def\half{{1\over 2}}
\def\threehalf{{3\over 2}}

\begin{document}
\thispagestyle{empty}
\setcounter{page}{0}

\begin{flushright}
hep-ph/9502201\\
MPI-PhT/95-9\\
February 1995
\end{flushright}
\bigskip
\bigskip
\bigskip
\bigskip
\bigskip

\begin{center}
{\Large\bf Coupling Constant Unification in Extended
SUSY Models}\\
\vspace{2em}
\large
Ralf Hempfling

{\it Max-Planck-Institut f\"ur Physik,
Werner-Heisenberg-Institut,}

{\it D--80805 M\"unchen, Germany}

\vspace{1.5ex}
{\it Emails:} {\tt hempf@iws186.mppmu.mpg.de}
\end{center}
\bigskip
\bigskip
\bigskip

\begin{abstract}
\noindent
The unification of gauge coupling constants
in the minimal supersymmetric model (MSSM)
is unaffected at the one-loop level by the inclusion of additional
mass-degenerate SU(5) multiplets.
Perturbativity puts an upper limit on the number of additional
fields. We analyse the evolution of the gauge coupling constants
in all models satisfying these criteria
using two-loop $\beta$ functions and including
low energy threshold effects.
We find that similarly to the minimal supersymmetric model (MSSM)
unification takes place within the theoretical and experimental errors.
The dominant proton decay mode is more suppressed in all extended models
as opposed to the MSSM due to
renormalization group effects.
However, the prediction for the bottom to $\tau$ mass ratio
becomes worse in all models under consideration.

\end{abstract}
\newpage

%\section{Introduction}

The prediction of the strong coupling constant,
$\alpha_s$, within the framework of the minimal supersymmetric model (MSSM)
assuming gauge coupling unification without any intermediate scale
%the existence of a unified gauge group
is in acceptable agreement with
experiment\cite{amal}.
In addition, the unification of $\tau$ and bottom Yukawa couplings
is quite promising\cite{dhr}.

Despite these successes of supersymmetric grand unified theories (SUSY GUTs)
based on SU(5)\cite{gutreview} there remain still some problems.
For example, the prediction for the down type
quarks from Yukawa unification
of the first two generations is off by a factor
$(m_d m_\mu)/(m_e m_s) = O(10)$.
Maybe the most sever challenge is the so-called
doublet/triplet problem\cite{doublettriplet}
of giving the colored Higgs triplet a mass of the order of $\mgut$ while
retaining the Higgs doublets responsible for the electro-weak symmetry
breaking at the electroweak scale.
There have been many attempt to try and solve these problems
all of which have one thing in common:
they require the introduction of new fields. Thus, an
extension of the particle content of the MSSM is inevitable
and one might ask whether some of these new particles are present in
the low energy effective theory.
This idea has already been explored in ref.~\cite{xsusy1,xsusy2}
and will be reconsidered here in view of the improved
experimental limits on the gauge coupling constants.
In addition, we will relax the constraint for perturbativity
and as a result obtain models that have not been considered before.

{}From the severity of the doublet/triplet problem we know
how hard it is to construct a model with a large mass hierarchy
between different SU(3)$\otimes$SU(2)$\otimes$U(1) members of
the same SU(5) multiplet. Therefore, we can expect that the most natural
extensions of the MSSM will come as complete SU(5) multiplets
in the low energy effective theory.
It is a simple exercise to show that all such extensions
will preserve gauge coupling unification at the one-loop level.

The renormalization group equations (RGEs) for these models
can be written as
\beqn
{d \alpha_i\over d t} =  \alpha_i^2[\beta_i + \alpha_j\beta_{i j} +
O(\alpha^2)]\,.
\label{rges}
\eeqn
Here, $t\equiv (2\pi)^{-1}\ln {(\rm scale)}$,
the indices $i,j = 1,2,3$ refer to the
U(1), SU(2) and SU(3) gauge group and summation over twice occurring
indices is assumed. Furthermore, the one-loop $\beta$ functions
for the gauge couplings are\cite{onelpgauge}
\beqn
\beta_i = \left(\matrix{0\cr-6\cr-9}\right)  +
  N_H    \left(\matrix{\threetenths\cr\half\cr0}\right)  +
\left(2 N_G + \beta^X\right)\,,
\label{rgeg}
\eeqn
where the three contributions to $\beta_i$ come from the gauge sector,
the Higgs doublets (in the MSSM the number of Higgs doublets, $N_H = 2$)
and the contribution of complete SU(5) multiplets.
The contributions of the gauge/gaugino sector to $\beta_i$ are
non-universal since some gauge bosons and their superpartners
acquire a mass via the Higgs mechanism
while others stay massless due to gauge invariance.
The contributions of the Higgs bosons are also non-universal because
the doublets are responsible for the electro-weak symmetry
breaking and should have a mass of the order of $100~\gev$
while the Higgs triplets have to acquire a mass at $\mgut$
in order to sufficiently suppress the rate for proton-decay.
This doublet/triplet splitting\cite{doublettriplet}
is solely motivated
by experiment and has no satisfying theoretical solution yet.
The last term which also contains the contribution of $N_G = 3$ generations
of quarks and leptons is universal for all three couplings.
(Note, that one family of quarks and leptons can be embedded in a
$\fivebar = d^c(\threebar,1,2/3)\oplus l(1,2,-1)$ and a
$10  = q(3,2,1/3)\oplus u^c(\threebar,1,-4/3)\oplus e^c(1,1,-2)$;
the numbers in brackets indicate the transformation properties under
the SU(3), SU(2) and U(1) gauge symmetries, respectively.)
The reason is that the inclusion of a full SU(5) multiplets with a
mass, $m$, does not break the SU(5) gauge symmetry and should yield a
universal contribution
to all three $\beta$ functions at any scale above $m$ at the one-loop level.
The contributions of the extensions of the MSSM can be written as
\beqn
\beta^X = \sum_\Phi T(\Phi)\,,
\label{betax}
\eeqn
where the sum is over all SU(5) multiplets $\Phi$.
The values of
$T(\Phi) \equiv d(\Phi) C_2(\Phi)/r$ are listed in Table~\ref{tphi}
for the four smallest representations of SU(5).
Here, $r = 24$ is the number of generators of SU(5) and $C_2(\Phi)$
[$d(\Phi)$]
is the quadratic Casimir operator [dimension]
of the SU(5) representation $\Phi$\cite{slansky}.
By imposing gauge coupling unification at $\mgut$ \ie
\beqn
\alpha_{\rm GUT}\equiv \alpha_1(\mgut)=\alpha_2(\mgut)=\alpha_3(\mgut)\,,
\label{bound}
\eeqn
and solving eq.~\ref{rges} to first order in perturbation theory we obtain
\beqn
t_0 \equiv {1\over 2\pi} \ln {\mgut\over \mz}
= {\alpha_1^{-1}(\mz) - \alpha_2^{-1}(\mz)\over \beta_1 - \beta_2} \simeq
5.3\,,\cr
\alpha_3^{-1}(\mz) = \alpha_2^{-1}(\mz) + t_0 (\beta_3 - \beta_2)  \simeq 8\,,
\label{solve}
\eeqn
where we have used
\beqn
\alpha_1(\mz) = {5\over 3} {\alpha_{\rm em}\over \cos^2 \theta_{\msbar}}\,,\cr
\alpha_2(\mz) =  {\alpha_{\rm em}\over \sin^2 \theta_{\msbar}}\,,\cr
\eeqn
and $\alpha^{-1}_{\rm em} = 127.9$
and $\sin^2 \theta_{\msbar} = 0.2319$\cite{pdg}
as the low energy input values.
The prediction of the strong coupling constant,
$\alpha_s(\mz) \equiv \alpha_3(\mz)
\simeq 0.125$ is in quite good agreement with the world average
$\alpha_s(\mz) = 0.117\pm0.005$~\cite{pdg}.
In deriving eq.~\ref{solve} we have assumed that there is
no intermediate scale but it also holds in the
case of a widely spread particle spectrum as long as the
members of the different SU(5) multiplets lie
close together.

Note, that the right hand side of eq.~\ref{solve}
is independent of $\beta^X$.
This means that any extension of the MSSM by full SU(5)
multiplets will maintain the property of gauge coupling unification
at one loop.
The unification scale $\mgut \simeq 2 \times 10^{16}~\gev$
remains also unchanged and for the unified gauge coupling we obtain
\beqn
\alpha^{-1} \simeq 24 - t_0 \beta^X\,.
\label{limitd}
\eeqn
By requiring that the right hand side of eq.~\ref{limitd} is larger than
zero we find $\beta^X \lsim 4.5$ but maybe models with $\beta^X = 5$
are still
acceptable due to higher order corrections or threshold corrections
{\it etc.} and shall be included into our considerations.
%\beqn
%\beta^X = \lsim 5
%\label{defdx}
%\eeqn

\begin{table}[t]
$$
\begin{array}{|c||c|c|c|c|}
\hline
\Phi    &     5,\fivebar     &   10,\tenbar      &     15,\fivteenbar    & 24
\\ \hline \hline
T(\Phi) & {1\over2} & {3\over2} & {7\over2} & 5
\\ \hline
\end{array}
$$
\caption{
The contribution of various SU(5) multiplets $\Phi$ to $\beta^X$.
}
\label{tphi}
\end{table}

In order to derive a viable model, we have to impose additional
constraints.
The cancellation of triangle anomalies implies that complex representations
only occur in pairs.
Thus, there are four types of extensions satisfying the above requirements
\begin{itemize}
\item $n$ additional pairs of $5$ and $\fivebar$, where $n = 1,2,3,4,5$,
\item one additional pair of $10$ and $\tenbar$,
\item $n$ additional pairs of $\fivebar$ and $10$ where $n = 1,2$,
\item one additional adjoint representation,
$24 = g(8,1,0)\oplus w(1,3,0) \oplus b(1,1,0)
\oplus x(3,2,5/3) \oplus \xbar(\threebar,2,-5/3)$.
\end{itemize}
Experimental lower limits on the additional
particle masses can be satisfied by
adding explicit  dirac or majorana mass terms into the superpotential
such as
\beqn
W \supset \mu_5 \fivebar 5, ~ \mu_{10} \tenbar 10, ~ \mu_{24} 24^2\,,
\eeqn
allowed by gauge invariance in models of type 1, 2 and 4.
Models of type 3 correspond to the MSSM with four or five generations.
For the invisible width of the $Z$ boson at LEP experiments we
know that the number of (almost) massless neutrinos is 3 and a mechanism
has to be introduced in order to give mass to the additional
neutrinos larger than about $\mz/2$. Also, a lower limit on the
mass of an additional lepton of $m_{\tau^\prime}\gsim \mz/2$ at LEP
and a lower limit on the
mass of an additional down type quark of $m_{b^\prime}>85~\gev$ at CDF
have been established\cite{pdg}.
This implies that the Yukawa couplings for the additional
fermions are bound from below since no explicit gauge
invariant mass term exists.
On the other hand, there is an upper limit on the masses
from the infra-red fixed-point behaviour of the Yukawa couplings
leaving only a very constrained region in parameter space.
The four generation model has been studied recently and found to be quite
constrained but could still be feasible if a right-handed
neutrino is introduced to raise the mass of the left-handed
neutrino above the experimental bounds\cite{gunion}.
The five generation model is even more
constrained and might already be ruled out by present data.
However, the model with additional $5, \fivebar, 10, \tenbar$ with the
possibility for a explicit dirac mass for all the
additional particles is still allowed.
It also does not require any additional fields
to generate mass for the unseen neutrinos.

\begin{table}[t]
$$
\begin{array}{|c||c|c|c|c|c|c|c|c|c|c|}
\hline
\beta^i_\phi&  l  & d^c  & e^c & u^c& q & b & w & g & x&\xbar
\\ \hline \hline
U(1) &-{3\over10}&-{2\over15}&-{6\over 5}&-{8\over15}&-{1\over30}
& 0 & 0 & 0 &-{5\over6} &-{5\over6}
\\ \hline
SU(2)&-{3\over 2}   & 0  & 0  & 0 &-{3\over 2} &
0 & -4 & 0 &-{3\over 2} &-{3\over 2}
\\ \hline
SU(3)& 0    &-{8\over 3}& 0 &-{8\over 3}&-{8\over 3}& 0& 0 &-6
&-{8\over 3}&-{8\over 3}
\\ \hline
\end{array}
$$
\caption{
The $\beta$ functions for the additional fields, $\phi$
}
\label{betaphi}
\end{table}

The assumption that all the members, $\phi$,
of one SU(5) multiplet, $\Phi$,  are mass
degenerate is protected by gauge invariance.
It acquires corrections below $\mgut$ where the gauge
symmetry is broken
%with a mass, $\mu_\Phi$, of the order of the Higgs mass
%parameter, $\mu$, at $\mgut$ is violated
through one-loop RG evolution
\beqn
{d \mu_{\phi}\over d t} = \mu_\phi\sum_i \alpha_i \beta_\phi^i\,,
\label{betamu}
\eeqn
where the $\beta^i_\phi$ are listed in Table.~\ref{betaphi}.
The solution of eq.~\ref{betamu} can be written as
\beqn
{\mu_{\phi}(\mu_\phi)\over \mugut} =
\Pi_i \left({\alpha_{\rm GUT}\over \alpha_i(\mu_\phi)}\right)
^{\beta_\phi^{i}/\beta_i}\,,
\label{rgemu}
\eeqn
where we have assumed that all possible additional
couplings of the superpotential
are much smaller than $\alpha_{\rm GUT}$ and can be neglected
in the evolution of the $\alpha_i$'s.
Furthermore, we assume that the complex (real) multiplets
are odd (even) under R-parity so that proton decay mediating interactions
are forbidden.
We see that the splitting between the various members of an SU(5)
multiplet is quite considerable and gives rise to
significant threshold corrections.
If we decouple the fields $\phi$ from the RGEs at $\mu_\phi$
we obtain the improved one-loop formulas
\beqn
\alpha_i^{-1}(\mz) = \alpha_{\rm GUT}^{-1}+t_0 \beta_i
- \Delta^{\rm MSSM}_i
-{1\over 2\pi}\sum_\phi N_\phi \ln{\mu_\phi\over \mz} \beta_{i,\phi}\,,
\label{oneloop}
\eeqn
where $\beta_{i,\phi}$ is the contribution of the field, $\phi$ to
$\beta_i$ listed in Table~\ref{tablebetaphi}\cite{massrge}. Furthermore,
\beqn
N_\phi = \left\{\matrix{N_5\cr N_{10} \cr N_{24}}\right\}~\hbox{for}
{}~\phi =   \left\{\matrix{l, d^c\cr e^c, u^c, q \cr b,w,g,x,\xbar}
\right.\,,
\eeqn
is the number of the fields $\phi$.
The MSSM threshold corrections, $\Delta^{\rm MSSM}_i$,
studied in ref.~\cite{mssmthresh} raise the predicted
value of $\alpha_s(\mz)$ by about 10\% and
spoil the success of the GUT prediction to some degree.

For the computation of $\Delta^{\rm MSSM}_i$ we have
assumed that the effects of SUSY breaking are
parameterized in the standard
fashion by including
explicit soft SUSY breaking terms assumed to be universal at
$\mgut$. Thus, at $\mgut$ we have only four independent
soft SUSY parameters:
the coefficients multiplying
the trilinear and quadratic terms of the superpotential, $A$ and $B$,
which are irrelevant for the mass spectrum and, thus,
for the value of the gauge couplings at
the leading log level, the mass for all spin 0 particles
$m_0$, and the gaugino mass parameter, $m_{1/2}$.
(We refrain from a more sophisticated treatment of the
SUSY threshold corrections since here we only care about
the changes of extended models with respect to the MSSM.)
{}From the last two parameters we can derive the
full low energy mass spectrum via RG evolution.

In deriving eq.~\ref{oneloop} we have also made the assumption
that the mass splitting between spin 0 and spin 1/2 components
due to soft SUSY breaking mass terms can be neglected.
This is motivated by the observation that
the Higgs mass parameter $\mu$ has to be of the order of the top
squark mass for correct radiative electro-weak symmetry
breaking. However, from eq.~\ref{rgemu} we see that
$\mu^2_\phi(\mu_\phi) \gg \mu^2(\mz)$
if $\phi$ carries color.
Also, $\mu^2(\mz)$ is additionally suppressed by the effect
of the large top Yukawa coupling, $\alpha_t$.
%a factor exp$(-3\int \alpha_t dt)$.
We will return to this point
in the context of radiative electro-weak symmetry breaking.

Additional effects on the unification of the gauge couplings
arise because the universality of $\beta^X$ in eq.~\ref{rgeg}
is violated at the two-loop level.
The two-loop $\beta$ functions of these models can easily be derived from
ref.~\cite{jones}
\beqn
\beta_{i j}^X = N_5 \left(\matrix{{7\over30} &{9\over10} &{16\over15} \cr
                                {3\over10} &{7\over 2} & 0 \cr
                                {2\over15} &         0 & {17\over 3}}\right)
       +N_{10} \left(\matrix{{23\over10} &{3\over10} &{72\over15} \cr
                                {1\over10} &{21\over 2} & 8 \cr
                                {9\over15} &         3 & 17}\right)
       +N_{24} \left(\matrix{{25\over3}     & 15         &{80\over3} \cr
                                5 &33 & 16 \cr
                                {10\over3} &         6 & {68\over3}}\right)\,.
\eeqn
Here we have not separated out the contributions
of the individual representations since the two-loop threshold corrections
analogous to the one-loop corrections
of eq.~\ref{oneloop} are negligible.

\begin{table}[t]
$$
\begin{array}{|c||c|c|c|c|c|c|c|c|c|c|}
\hline
\beta_{i,\phi}& l & d^c & e^c  & u^c  & q  & b & w & g & x & \xbar
\\ \hline \hline
U(1) & {3\over 10}& {1\over 5}& {3\over 5}& {4\over 5}& {1\over 10}& 0 & 0 & 0
&{5\over2} & {5\over2}
\\ \hline
SU(2)& {1\over 2} & 0         & 0         & 0         & {3\over 2} & 0 & 2 & 0
 & {3\over2}& {3\over2}
\\ \hline
SU(3)& 0          & {1\over 2}& 0         & {1\over 2}& 1 & 0 & 0 & 3 & 1 & 1
\\ \hline
\end{array}
$$
\caption{
The one-loop $\beta$ functions for the
SU(3)$\otimes$SU(2)$\otimes$U(1) gauge couplings due to the
individual representations $\phi$.
Note that by summing over a complete SU(5) multiplet we recover
the universal coefficients of Table~1
}
\label{tablebetaphi}
\end{table}

In minimal SU(5) SUSY-GUTs the down-type quark fields $d$ and
the left-handed lepton fields $l$ are embedded in
one representation and as a result the
$\tau$ and bottom Yukawa couplings
are unified at $\mgut$.
In the limit of negligible $\tau$ and bottom Yukawa couplings
(\ie\ $\alpha_\tau, \alpha_b \ll\alpha_{\rm GUT}$) we can
write in the one-loop approximation
\beqn
{m_b(\mz)\over m_\tau(\mz)} =
\Pi_i \left({\alpha_{\rm GUT}\over
\alpha_i(\mu_\phi)}\right)^{(\beta_b^{i}-\beta_\tau^i)/\beta_i}
{\rm exp}\left(-{1\over 2}\int_0^{t_0} \alpha_t d t\right)\,,
\label{rgebt}
\eeqn
where $\beta^i_b = (7/18,3/2,8/3)$ and $\beta^i_\tau = (3/2,3/2,0)$.
It is easy to see that $m_b/m_\tau$ increases with
$\alpha_{\rm GUT}$ and hence also with $\beta^X$.
This increase can be compensated by an increase in the top
Yukawa coupling, $\alpha_t$.
However, $\alpha_t$ quickly approaches its IR fixed-point
$\alpha_t \simeq (8/9)\alpha_s$ (near $\mgut$:
$\alpha_t \simeq (44/27)\alpha_{\rm GUT}$).
Hence, any significant increase of the value of the integral
in eq.~\ref{rgebt} can only come from the integration close to $\mgut$
and requires unperturbatively large values of $\alpha_t(\mgut)$.
Thus, in our numerical work we have used
$\alpha_t(\mgut) = \alpha_{\rm GUT}$ in order to obtain
a natural upper limit for the integral in eq.~\ref{rgebt}
and, hence, a natural lower bound on $m_b/m_\tau$ for a
particular model assuming $\tau$-bottom Yukawa unification.

Another important constraint on SUSY-GUT models
comes from non-observation of proton-decay.
In the MSSM the dominant decay proceeds via
$p \to K+\nu$\cite{pdecay}
by dressing of the dimension 5 operators obtained from
the non-renormalizable term of the superpotential
\beqn
W_{NR}    = \Gamma (q q)(q l)\,,
\eeqn
where the SU(2) indices are contracted inside the brackets
and flavor and color indices are omitted.
At the one-loop level we find
\beqn
\Gamma \propto {1\over \mgut}m_u m_d
\Pi_i\left({\alpha_i(\mz)\over \alpha_{\rm GUT}}\right)
^{(\beta^i_u+\beta^i_d
-\threehalf \gamma^i_q - \half \gamma^i_l)/\beta_i}\,.
\eeqn
Here, we have defined
$\beta^i_d = \beta^i_b$,
$\beta^i_u = (13/18,3/2,8/3)$,
$\gamma^i_q = (1/10,3/2,8/3)$, and
$\gamma^i_l = (1/2,3/2,0)$.
Of course, a precise determination of the value of $\Gamma$
is problematic without a selfconsistent model for
the origin of the Yukawa couplings for all three generations.
However, we do not worry about any overall factors since
we are only interested in the ratio $\Gamma^X/\Gamma^{\rm MSSM}$.

\begin{table}[t]
$$
\begin{array}{|c||c|c|c||c|c||c|c||c|c||c|c||c|c|}
\hline
{}&N_5 & N_{10} & N_{24}
    & \multicolumn{2}{c||}{r}
    & \multicolumn{2}{c||}{ \Delta \alpha_s/\alpha_s {\rm in \%} }
    & \multicolumn{2}{c||}{\alpha_{\rm GUT}}
    & \multicolumn{2}{c||}{ m_b/m_\tau }
    & \multicolumn{2}{c|}{ \Gamma^X/\Gamma^{\rm MSSM}}
\\ \hline \hline
{\rm MSSM}&
   0&0&0&  .72& 1.00&  -6.4&   0.0&   .040&   .042&  1.80&  1.87&   1.0&   1.0
\\ \hline
1& 2&0&0&  .75& 1.11&  -4.8&   -.1&   .050&   .053&  1.88&  1.96&   .86&   .79
\\ \hline
2& 4&0&0&  .79& 1.29&  -2.7&   -.1&   .066&   .073&  1.99&  2.09&   .70&   .57
\\ \hline
3& 6&0&0&  .85& 1.59&   .43&   -.1&   .096&   .118&  2.15&  2.30&   .51&   .34
\\ \hline
4& 8&0&0&  .97& 2.16&   6.3&  -1.8&   .175&   .333&  2.44&  2.65&   .29&   .13
\\ \hline
5& 1&1&0&  .71& 1.63&  -6.1&   1.3&   .070&   .081&  2.08&  2.31&   .71&   .38
\\ \hline
6& 3&1&0&  .76& 2.23&  -3.7&   1.5&   .104&   .149&  2.28&  2.64&   .51&   .18
\\ \hline
7& 5&1&0&  .83& 4.20&   .73&  -2.1&   .203&   1.68&  2.64&  3.36&   .27&   .03
\\ \hline
8& 0&2&0&  .48& 1.57&  -9.3&   -.2&   .088&   .118&  2.01&  2.28&   1.0&   .35
\\ \hline
9& 2&2&0&  .46& 2.16&  -7.6&  -1.6&   .144&   .332&  2.18&  2.62&   .79&   .14
\\ \hline
10& 0&0&1&  .49&  .72&  -3.3&  26.&   .280&  2.38&  2.32&  3.26&    .52&   .18
\\ \hline
\end{array}
$$
\caption{
The one-loop (left column) and two-loop (right column) results for
the ratio, $r = \mgut^{X}/\mgut^{\rm MSSM}$,
$\Delta\alpha_s/\alpha_s$ [in \%],
$\alpha_{\rm GUT}$,
the ratio $m_b/m_\tau$.
and the ratio $\Gamma^X/\Gamma^{\rm MSSM}$.
We have chosen $m_0(\mgut) = 200~\gev$,
$\mu_\phi(\mgut) = 1~\tev$ for all additional fields, $\phi$,
and $\alpha_t(\mgut) = \alpha_{\rm GUT}$.
Furthermore, we have fixed $m_{1/2}$ such that
$M_{\tilde g}(M_{\tilde g}) = 200~\gev$.
}
\label{result}
\end{table}

In Table~\ref{result} we have summarized our results for
the MSSM and 8 extended models characterized by
$N_5$, $N_{10}$, and $N_{24}$.
We have chosen the somewhat large values of
$\mu_\phi = 1~\tev$ for all additional fields $\phi$
in order to exploit as many models as possible.
The parameter independent lower limit on the gluino mass of
$M_{\tilde g} > 100~\gev$ has been established from direct
particle search at CDF\cite{refgluino}.
However, stronger limits can be
derived from the chargino/neutralino search by imposing
GUT constraints and we chose
$M_{\tilde g} = 200~\gev$ in order to safely
avoid all the present bounds.

The first row corresponds to the MSSM where we have
$\alpha_s^{\rm MSSM} = 0.124$ and
$\mgut^{\rm MSSM} = 2.3\times 10^{16}~\gev$
for our choice of parameters.
In the different columns we present the result for the
unification scale divided by the two-loop MSSM value and denoted by $r$,
the relative change in the prediction of $\alpha_s$
with respect to the two-loop MSSM value,
the unified gauge coupling, $\alpha_{\rm GUT}$, and the ratio
of $m_b(\mz)$ to $m_\tau(\mz)$.
The two values correspond to the results obtained by using one-loop
and two-loop $\beta$ functions.
We see that already for the MSSM the
value of $m_b/m_\tau$ for $\alpha_t(\mgut) = \alpha_{\rm GUT}$
is slightly above its experimental value of $m_b/m_\tau \simeq 1.6$
but can still be brought in agreement with
experiment by a modest increase of $\alpha_t$
or by choosing $\alpha_b = O(\alpha_t)$\cite{dhr}.
The situation becomes more problematic in all
extended models where this ratio increases even more.
%and cannot be reconciled with experiment in all models with non-perturbative
%gauge unification (ie. $\beta_3 > 0$)
%by a mere increase of $\alpha_t$.

Furthermore, we see that the models 4, 7, 9 and 10 become
non-perturbative at $\mgut$.
This scenario of non-perturbative unification
was already advocated in ref.~\cite{nonpert1} in non-SUSY models
and in ref.~\cite{nonpert2} extended to SUSY models
as being particularly attractive.
The reason is that the dependence of $\alpha_i$ on
$\alpha_{\rm GUT}$ in eq.~\ref{oneloop} vanishes in the
large $\alpha_{\rm GUT}$ limit.
In the presence of explicit mass terms this argument becomes more
complicated.

{}From eq.~\ref{rgemu} we see that the low energy particle
spectrum becomes non-predictive when $\alpha_{\rm GUT}$ vanishes.
However, we will still be able to use perturbation theory
to predict certain trends in these models.
For example, at the one-loop level the gluino mass satisfies
\beqn
M_{\tilde g}(M_{\tilde g}) = {\alpha_s(M_{\tilde g})\over
\alpha_{\rm GUT}} m_{1/2}\,,
\eeqn
and thus we find
$M_{\tilde g}(M_{\tilde g})\ll m_{1/2}$ in all models
with non-perturbative gauge unification.
On the other hand, we have for the squark masses
\beqn
M^2_{\tilde Q} = C m_{1/2}^2 + ...\,,
\eeqn
where $C \simeq 6- 0.8 \beta^X\gsim 2$ and we have
dropped irrelevant terms.
% in the MSSM and this factor increases in all extended models.
Thus, by imposing
$M_{\tilde g}(M_{\tilde g}) = 200~\gev$ we find typically that
all the squark masses have to be in the TeV region or higher
depending on $\alpha_{\rm GUT}$.
Such a large hierarchy between the squark masses and
the electro-weak scale requires fine-tuning and is problematic
as we will see in the following.
In the limit of small gluino mass
the soft SUSY breaking
top squark mass parameters, $M_{\tilde t_P}^2$ ($P = L,R$)
and the Higgs mass parameter, $M_{H_2}$,
are closely related by the solution to their RG equations
\beqn
{\Sigma(\mz)\over \Sigma(\mgut)}=
\exp\left(-6\int_0^{t_0} \alpha_t d t\right)\,,
\label{rgesig}
\eeqn
where we have defined
$\Sigma \equiv M_{\tilde t_L}^2 + M_{\tilde t_R}^2 + M_{H_2}^2$.
In the limit of large $\alpha_t$ the right hand side of eq.~\ref{rgesig}
vanishes and we find
\beqn
\abs{M_{H_2}^2} \simeq  M_{\tilde t_L}^2 + M_{\tilde t_R}^2
\gsim (1~\tev)^2\,.
\eeqn
The correct electro-weak symmetry breaking then implies that
we have to
fine-tune $\mu$ and $M_{H_2}^2$ over several orders of magnitude in order
to satisfy
\beqn
M_{H_2}^2 + \mu^2 = 0(\mz^2)\,.
\eeqn
This heavy SUSY particle spectrum and the associated
fine-tuning is a generic problem in all models with non-perturbative
unification over the entire parameter space.

It is interesting to note that model~10 automatically provides a solution
to this problem. Here, the gluino can
acquire a mass via mixing with the octet of the $24$ dimensional
representation\cite{xsusy1} and as a result the mass parameter
$M_{\tilde g}(M_{\tilde g})$ can be much smaller than the lower bound on
the gluino mass.
Such a dirac mass term can be compatible with $R$ symmetry\cite{rsymm1}
and was already advocated in ref.~\cite{rsymm2} as the origin of
the gluino mass.
Furthermore, a model with an approximate $R$ symmetry
naturally yields large value of Higgs vacuum expectation values,
$\tan\beta$, which explains the large ratio of $m_t$ to $m_b$.
In the MSSM this can only be achieved by fine-tuning\cite{finetuning}.

In order to see how far we can trust
perturbation theory in the case of
non-perturbative unification we will modify the boundary condition
of eq.~\ref{bound} for model~10.
In Table~\ref{modibound} we have listed the
prediction for $\alpha_{\rm em}(\mz)$, $\sin^2 \theta_{\msbar}$,
and $\alpha_s(\mz)$. We have used $\mgut = 1.6\times 10^{16}$
and $\alpha_{\rm GUT} = 3.2$.

\begin{table}[t]
$$
\begin{array}{|c|c|c||c|c|c|}
\hline
\alpha_1/\alpha_{\rm GUT}&
\alpha_2/\alpha_{\rm GUT}&
\alpha_3/\alpha_{\rm GUT}&
\alpha_{\rm em}^{-1}&
\sin^2 \theta_{\msbar}&
\alpha_s
\\ \hline \hline
1&1&1&  127.9 &  .232 &  .156
\\ \hline
0.1&1&1&132.5 &  .224 &  .158
\\ \hline
1&0.1&1&130.2 &  .246 &  .161
\\ \hline
1&1&0.1&126.3 &  .229 &  .112
\\ \hline
10&1&1& 127.6 &  .232 &  .155
\\ \hline
1&10&1& 127.9 &  .231 &  .154
\\ \hline
1&1&10& 128.7 &  .234 &  .155
\\ \hline
\end{array}
$$
\caption{
The prediction of the low energy gauge coupling constants
for various boundary conditions at $\mgut$ in model~10
}
\label{modibound}
\end{table}

We find indeed that if $\mgut$ is close to a Landau pole
any potentially large threshold corrections get washed out
via RG evolution into the perturbative region below
$\mgut$. We see that changes of the GUT input parameters
by a factor of 10 only changes the low energy values by typically
a few \%\ (the exception is the case where
$\alpha_3 \\l \alpha_1, \alpha_2$ because here, the splitting
of the $24$ due to eq.~\ref{rgemu} is only mild).
As a result, a theory that is non-perturbative at
$\mgut$ might still allow for reliable low energy predictions.
In fact, in model~10 the three gauge couplings
are completely fixed by only one input parameter, $\mgut$,
rather than two as in the MSSM.
However, it is somewhat surprising that the
predictability of the model
is not so much limited by non-perturbative effects at $\mgut$
but rather by the low energy threshold corrections in
eq.~\ref{oneloop}.
The reason is that the particle spectrum derived in eq.~\ref{rgemu}
becomes non-predictive in the large $\alpha_{\rm GUT}$ limit
and can only provide us with a qualitative understanding.
In general, we
expect colored particles to be much heavier than the
electro-weak scale and to decouple at a higher scale.
This predicts an decrease of
$\alpha_3^{-1}$ due to eq.~\ref{oneloop} which
for our particular choice of parameters
and using our two-loop RG approach
results in an overestimate of $\alpha_s$ by
about 30\%.
This number should not be interpreted as a prediction but rather
as an indication for how the unpredictability
of the particle spectrum feeds into the prediction of
the coupling constants.

In summary, we have investigated 10 extensions of the MSSM
with additional SU(5) multiplets. We find that in all models
where $\alpha_{\rm GUT}<1$
the prediction of $\alpha_s$ only changes by a few \%.
Furthermore, the ratio
$m_b/m_\tau$ increase and the proton decay rate decreases
with increasing $\alpha_{\rm GUT}$.
We point out that models with non-perturbative unification
predict in general a large squark to gluino mass ratio which leads
to a fine-tuning problem once we impose the experimental lower limit
on the gluino mass.
The exception is the model with an additional 24 dimensional
representation where the gluino can have a dirac mass.
However, SUSY threshold corrections to $\alpha_s$ are
positive and large due to large multiplet splitting
and in general raise $\alpha_s$
significantly above the  experimental limit.


\newpage
\begin{thebibliography}{References:}

\bibitem{amal} U. Amaldi, W. de Boer and H. F\"urstenau,
\PLB {260}{443}{1991};
J. Ellis, S. Kelley and D.V. Nanopoulos, \PLB {260}{131}{1991};
P. Langacker and M.X. Lou, \PRD {44}{817}{1992}.

\bibitem{dhr} S. Dimopoulos, L.J. Hall and S. Raby, \PRL {68}{1984}{1992};
\PRD {45}{4192}{1992};
V. Barger, M.S. Berger and P. Ohmann, \PRD {47}{1093}{1993};
M. Carena, S. Pokorski and C.E.M. Wagner, \NPB {406}{59}{1993};
P. Langacker and N. Polonsky, \PRD{47}{1093}{1993};
W.A. Bardeen, M. Carena, S. Pokorski and C.E.M. Wagner, \PLB{320}{110}{1994}.

\bibitem{gutreview} for a review of GUT theories see:
R. Mohapatra, {\sl Unification and Supersymmetry,}
2nd ed. (Springer Verlag, Berlin, 1992);
G.G. Ross, {\sl Grand Unified Theories}
(Addison-Wesley, Reading, MA, 1984).

\bibitem{doublettriplet} S. Dimopoulos and H. Georgi,
\sl Nucl. Phys. \bf B193\rm , 150 (1981).

\bibitem{xsusy1}
L. Ib\'a${\rm\tilde n}$ez, \PLB{126}{196}{1983};
(E) {\bf B130}, 463 (1983).

\bibitem{xsusy2}
J.E. Bj\"orkman and D.R.T. Jones, \NPB{259}{533}{1985}.

\bibitem{onelpgauge} M.B. Einhorn and D.R.T. Jones, \NPB{196}{475}{1981}.

\bibitem{slansky} For a review of group theory for
unified model building, see: R. Slansky, \PREP{79}{1}{1981}.

\bibitem{pdg} Particle Data Group, L. Montanet \etal, \PRD{50}{1173}{1994}.

\bibitem{gunion} J.F. Gunion, D.W. McKay and H. Pois, \PLB{334}{339}{1994}.

\bibitem{massrge} K. Inoue, A. Kakuto, H. Komatsu and S. Takeshita,
\PTP{67}{1889}{1982};
J.P. Derendinger and C.A. Savoy, \NPB{253}{285}{1985};
N.K. Falck, \ZPC {30}{247}{1986}.

\bibitem{mssmthresh} G.G. Ross and R.G. Roberts, \NPB{377}{571}{1992};
P. Langacker and N. Polonski, \PRD{47}{4029}{1993};
M. Carena, S. Pokorski and C.E.M. Wagner, \NPB{}{}{1993}.

\bibitem{jones} D.R.T. Jones, \PRD{25}{581}{1982}.

\bibitem{pdecay} J. Hisano, H. Murayama and T. Yanagida,
\NPB{402}{46}{1993} and references therein.

\bibitem{refgluino} F. Abe {\it etal}
   [CDF Collaboration], \PRL{69}{3439}{1992}.

\bibitem{nonpert1} L. Maiani, G. Parisi and R. Petronzio,
\NPB{136}{115}{1978};

\bibitem{nonpert2} N. Cabibbo and G.R. Farrar,
\PLB{110}{107}{1982}.

\bibitem{rsymm1} A. Salam and J. Strathdee, \NPB{87}{85}{1995};
P. Fayet, \NPB{90}{104}{1975}.

\bibitem{rsymm2} L.J. Hall and L. Randall, \NPB{352}{289}{1991}.


\bibitem{finetuning}
A.E. Nelson and L. Randall, \PLB{316}{516}{1993};
L.J. Hall, R. Rattazzi and U. Sarid,
LBL preprint LBL-33997 (1993), \sl Phys. Rev. \bf D, \rm to be published;
R. Hempfling, DESY preprint, DESY 94-078.




\end{thebibliography}
\end{document}